\documentclass[prd,aps,twocolumn,showpacs,nofootinbib]{revtex4-1}

\usepackage{amsfonts}
\usepackage[dvipdfm,colorlinks=true,citecolor=blue, linkcolor=blue,urlcolor=blue]{hyperref}
\usepackage{color,graphicx,amsfonts,multirow}
\usepackage{amsmath}

\begin{document}

\title{Photoproduction of doubly heavy baryon at the LHeC}

\author{Huan-Yu Bi$^{1}$}
\email{bihy@mail.ustc.edu.cn}
\author{Ren-You Zhang$^{1}$}
\email{zhangry@ustc.edu.cn}
\author{Xing-Gang Wu$^{2}$}
\email{wuxg@cqu.edu.cn}
\author{Wen-Gan Ma$^{1}$}
\email{mawg@ustc.edu.cn}
\author{Xiao-Zhou Li$^{1}$}
\email{lixiaozhou@ustc.edu.cn}
\author{Samuel Owusu$^{1}$}
\email{samuel@mail.ustc.edu.cn}

\address{$^1$ Department of Modern Physics, University of Science and Technology of China, Hefei, Anhui 230026, P.R. China}
\address{$^2$ Department of Physics, Chongqing University, Chongqing 401331, P.R. China}


\begin{abstract}

The photoproduction of doubly heavy baryon, $\Xi_{cc}$, $\Xi_{bc}$, and $\Xi_{bb}$, is predicted within the nonrelativistic QCD at the Large Hadron Electron Collider (LHeC). The $\Xi_{QQ'}$ production via the photon-gluon fusing channel $\gamma + g \to \langle{QQ'}\rangle[n] +\bar{Q} +\bar{Q'}$ and the extrinsic heavy quark channel $\gamma + Q \to \langle{QQ'}\rangle[n]+\bar{Q'}$ have been considered, where $Q$ or $Q'$ stand for heavy $c$ or $b$ quark and $\langle{QQ'}\rangle[n]$ stands for a $QQ'$ diquark with given spin- and color- configurations $[n]$. The diquark shall fragmentate into $\Xi_{QQ'}$ baryon with high probability. For $\Xi_{cc}$ and $\Xi_{bb}$ production, $[n]$ equals $[^1S_0]_{\textbf{6}}$ (in configurations spin-singlet $^1S_0$-wave and color-sextuplet $\textbf{6}$) or $[^3S_1]_{\bar{\textbf{3}}}$ (in configurations spin-triplet $^3S_1$-wave and color-antitriplet $\bar{\textbf{3}}$) and for $\Xi_{bc}$ production, $[n]$ equals $[^1S_0]_{\bar{\textbf{3}}}$, $[^1S_0]_{{\textbf{6}}}$, $[^3S_1]_{\bar{\textbf{3}}}$, or $[^3S_1]_{\textbf{6}}$. A detailed comparison of those channels and configurations on total and differential cross sections, together with their uncertainties, is presented. We find the dominant contributions for $\Xi_{QQ'}$ production are from extrinsic heavy quark channel and for $[^3S_1]_{\bar{\textbf{3}}}$ configuration, while other diquark states can also provide sizable contributions to the $\Xi_{QQ'}$ production. As a combination of all the mentioned channels and configurations and by taking $m_c=1.80\pm0.10$ GeV and $m_b=5.1\pm0.20$ GeV, we observe that $(6.70^{+1.40}_{-1.10})\times 10^{3}$ $\Xi_{bb}$, $(8.81^{+3.08}_{-2.20})\times 10^{6}$ $\Xi_{cc}$, and $(4.63^{+1.21}_{-0.92})\times 10^{5}$ $\Xi_{bc}$ events can be generated at the LHeC in one operation year with colliding energy $\sqrt{S}=1.30$ TeV and luminosity $ {\mathcal L}\simeq 10^{33}$ ${\rm cm}^{-2} {\rm s}^{-1}$. Thus, the LHeC shall help us to get more information about the doubly heavy baryon, especially for $\Xi_{cc}$ and $\Xi_{bc}$.

\end{abstract}

\maketitle

\section{Introduction}

The heavy-quark mass provides a natural hard scale, so the processes involving heavy quarks are perturbative QCD (pQCD) calculable by applying proper factorization theories. Among them, the nonrelativistic quantum chromodynamics (NRQCD)~\cite{NRQCD} provides a powerful approach to study the properties of doubly heavy hadrons, since their heavy constituent quarks move nonrelativistically in the bound system \cite{QWGC1, QWGC2}. Recently, we have studied the photoproduction of $B_c$ meson at the Large Hadron Electron Collider (LHeC)~\cite{LHeC} within the framework of NRQCD, and it was found that sizable amounts of $B_c$ meson events can be generated \cite{lhecbc}. As a step forward, we shall investigate whether sizable amounts of doubly heavy baryon events can also be produced at the LHeC. If so, to compare with future possible data, it could be inversely treated as a platform for testing the effectiveness of NRQCD.

For doubly heavy baryons $\Xi_{cc}$, $\Xi_{bc}$, and $\Xi_{bb}$, only $\Xi_{cc}$ was observed in the fixed target experiment by the SELEX collaboration~\cite{selex1, selex2, selex3}, which however has not been confirmed by other experiments so far. Here and henceforth, for simplicity, we use $\Xi_{QQ'}$ to denote the doubly heavy baryon $\Xi_{QQ'q}$, where $q$ stands for a light quark $u$, $d$, or $s$, respectively. At present, most of the predictions on the production rate and decay width are well underestimated compared to the SELEX measurements \cite{prodxiqq1, xicc1, xicc2, xicc3, xicc4, xicc5, xicc6, xicc7, fpro1, fpro3}. The unexpected high production rate of $\Xi_{cc}$ baryons may be due to the kinematics features of the SELEX experiment, and could not be described by the production mechanism only~\cite{Koshkarev:2016rci}. Thus, to understand the properties of doubly heavy baryon better, experimentally, we need to accumulate more events to reduce the statistical error; and theoretically, it is important to make a systematic research on various production mechanism of doubly heavy baryon at different experimental platforms.

A dedicated generator GENXICC \cite{GENXICC1, GENXICC11, GENXICC2} was developed and has been applied to study the production of doubly heavy baryon at the hadronic colliders \cite{prodxiqq2, prodxiqq3}. In 2013, the LHCb collaboration at the LHC performed their first search of $\Xi_{cc}$, but no significant signal has been found~\cite{Aaij:2013voa}. Future searches at the LHC with improved trigger conditions and larger data samples should improve the sensitivity significantly. In addition, series predictions on the production of doubly heavy baryon at the high luminosity electron-electron colliders, such as the $Z$-factory \cite{zfactory} and the International Linear Collider (ILC)~\cite{Djouadi:2007ik}, have also been performed. Sizable amounts of doubly heavy baryon events are expected to be generated in those platforms \cite{eexiqq1, eexiqq2, eexiqq3, ggxiqq}.

In addition, the hadron-lepton collider may also be a possible machine to probe the properties of the doubly heavy baryons. As clarified in Refs.\cite{LHeC, lhecbc}, the photoproduction mechanism dominates the production of $c/b$-quark at the LHeC, and the doubly heavy baryon can thus be mainly generated via the photoproduction channels $\gamma + g \to \Xi_{QQ'}+\bar{Q} +\bar{Q'}$ and $\gamma + Q \to \Xi_{QQ'}+\bar{Q'}$.

\begin{figure}[tb]
\begin{center}
\includegraphics[width=0.45\textwidth]{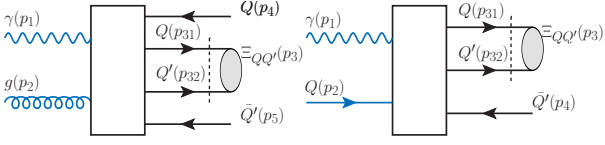}
\caption{Schematic diagrams for the photoproduction of $\Xi_{QQ'}$ at the LHeC. The box stands for the hard interaction kernel of $Q\bar{Q}$ ($Q'\bar{Q'}$)-pair production. } \label{gg}
\end{center}
\end{figure}

Schematic diagrams for the photoproduction of $\Xi_{QQ'}$ at the LHeC are shown in Fig.~\ref{gg}. The photoproduction of $\Xi_{QQ'}$ can be divided into three steps. Taking $\gamma+g$ channel as an example, the first step is the production of $Q\bar{Q}$ and $Q'\bar{Q'}$ pairs, where the heavy quarks $Q$ and $Q'$ are required to be in the color- and spin-configuration $[n]$; The second step is the $(QQ')[n]$ pair fuses into a binding diquark  $\langle QQ'\rangle[n]$ with certain probability; The third step is the evolution of the diquark into a doubly heavy baryon $\Xi_{QQ'}$ by grabbing a light quark from the ``vacuo" or emitting/grabbing a suitable number of gluons. The first step is perturbatively calculable since the gluon should be hard enough to generate the heavy quark-antiquark pair. For the second step, the transition probability can be described by a nonperturbative NRQCD matrix element. We use $h_{\textbf{6}}$ and $h_{\bar{\textbf{3}}}$ to stand for the matrix elements of the production of a color-sextuplet (${\textbf{6}}$) and a color-antitriplet (${\bar{\textbf{3}}}$) diquark, respectively \footnote{Here, we don't distinguish the matrix elements of $^1S_0$ and  $^3S_1$ states, since the spin-splitting effect is small \cite{prodxiqq2, eexiqq1}.}. For the third step, one usually assumes the efficiency of evolution from a $\langle QQ'\rangle[n]$ diquark to a doubly heavy baryon $\Xi_{QQ'}$ is $100\%$, referring as the ``direct evolution," Reference~\cite{ggxiqq} has studied the evolution through direct evolution as well as ``evolution via fragmentation" in which the fragmentation function has been taken into account. The authors there have found that the direct evolution is of high precision and is sufficient enough for studying the production of doubly heavy baryon, and thus we adopt the direct evolution in our calculation.

Since the predicted production rate of $\Xi_{cc}$ are much smaller than the SELEX measurements, the authors of Refs.\cite{xicc6, xicc7, GENXICC1} suggested to take into account the extrinsic and intrinsic charm production mechanism such that to shrink the gaps between theoretical and experimental predictions. It is noted that the intrinsic charm's contribution to the cross section of $\gamma + c$ channel is less than $0.1\%$ even if the density of intrinsic $c$-component in proton is up to $1\%$ \cite{intr1, intr2}. According to our experience on the $B_c$ meson photoproduction and following the suggestion given in Refs.\cite{xicc6, GENXICC1}, we shall concentrate our attention on the channels $\gamma + g \to \Xi_{QQ'} +\bar{Q} +\bar{Q'}$ and $\gamma + Q \to \Xi_{QQ'}+\bar{Q'}$, in which the intermediate diquark $\langle{QQ'}\rangle[n]$ is $\langle cc/bb\rangle[^1S_0]_{\textbf{6}}$, $\langle cc/bb \rangle[^3S_1]_{\bar{\textbf{3}}}$, $\langle bc\rangle[^1S_0]_{{\bar{\textbf{3}}}/{\textbf{6}}}$, and $\langle bc \rangle[^3S_1]_{{\bar{\textbf{3}}}/{\textbf{6}}}$, respectively. Other diquark configurations $\langle cc/bb \rangle[^3S_1]_{\textbf{6}}$ and $\langle cc/bb \rangle[^1S_0]_{\bar{\textbf{3}}}$ are forbidden due to the Fermi-Dirac statistics for identical particles.

The rest of the paper is organized as follows. In Sec.~\ref{sec2}, we present the formulations for dealing with the subprocesses $\gamma + g \to \Xi_{QQ'}+\bar{Q} +\bar{Q'}$ and $\gamma + Q \to \Xi_{QQ'}+\bar{Q'}$ in detail. Numerical results, theoretical uncertainties and discussions are given in Sec.~\ref{sec3}. Finally, a brief summary is given in Sec~\ref{sec4}.

\section{Calculation technology}\label{sec2}

\subsection{Basic formulas}

For the photoproduction mechanism, the initial photon is emitted from the electron, which can be described by the Weizs$\ddot{\rm a}$cker-Williams approximation (WWA)~\cite{wwa1, wwa2, wwa3}. When considering the extrinsic heavy-quark mechanism, we should pay special attention to avoid the ``double counting" problem between the $\gamma +g$ and the extrinsic $\gamma + Q$ channels. A proper approach to deal with the extrinsic heavy quark is to employ the general-mass variable-flavor-number scheme (GM-VFNs)~\cite{GMVFN1, GMVFN22, GMVFN2, GMVFN3, GMVFN4}. According to the pQCD factorization theorem, the cross section for the photoproduction of $\Xi_{QQ'}$ within the GM-VFNs scheme can be written as
\begin{widetext}
\begin{eqnarray}
&&d\sigma( e^- +P\to \Xi_{QQ'} + X) \nonumber \\
=
&&\sum_{[n]}
\langle{\cal O}^{\langle QQ' \rangle}[n]\rangle
 \bigg \{    f_{\gamma/e^-}(x_1)  f_{g/P}(x_2,\mu_f)\otimes d\hat{\sigma}({\gamma+g}\to (QQ')[n] + X)  \nonumber \\
&&~+ f_{\gamma/e^-}(x_1)  \Big[\Big( f_{Q/P}(x_2, \mu_f) - f_{Q/P}(x_2, \mu_f)_{\rm SUB} \Big)\otimes
  d\hat{\sigma}({\gamma+Q}\to (QQ')[n] + X)+ Q\longleftrightarrow Q' \Big]\frac{1}{1+\delta_{QQ'}} \bigg \},~~~
\label{eq1}
\end{eqnarray}
\end{widetext}
where $f_{\gamma/e^-}(x_1)$ is the WWA photon density function, $f_{i/P}(x_2,\mu_f)$ is the parton distribution function (PDF) of parton $i$ inside a proton $P$, $\mu_f$ is the factorization scale, and $d\hat{\sigma}({\gamma+i}\to (QQ')[n] + X)$ is the hard cross section for the partonic process $\gamma+i\to (QQ')[n]+X$. $\langle{\cal O}^{\langle QQ' \rangle}[n]\rangle$ is the nonperturbative matrix element which represents the transition probability from the $(QQ')[n]$-quark pair to the desired baryon $\Xi_{QQ'}$. $\delta_{QQ'}=1 (0)$ for $Q=Q'$ ($Q\neq Q'$). Since we adopt the direct evolution scheme, we have $\langle{\cal O}^{\langle QQ' \rangle}[n]\rangle$ = $h_{\bar{\textbf{3}}}$ or $h_{\textbf{6}}$, respectively.

The photon density function depicted by the WWA is expressed as~\cite{wwa1, wwa2, wwa3}
\begin{eqnarray}
f_{\gamma/e^-}(x)= \frac{\alpha}{2 \pi} \bigg [ \frac{1+(1-x)^2}{x} {\rm ln} \frac{Q^2_{\rm max}}{Q^2_{\rm min}} +\nonumber \\
 2 m_{e}^2 x \bigg (\frac{1}{Q^2_{\rm max}} -\frac{1}{Q^2_{\rm min}}\bigg) \bigg ],\label{wwaf}
\end{eqnarray}
where $x={E_{\gamma}}/{E_{e}}$, $E_{\gamma}$ and $E_e$ are photon and electron energies. $\alpha$ is the fine structure constant and $m_{e}$ is the electron mass. $Q^2_{\rm min}$ and $Q^2_{\rm max}$ are given by
\begin{equation}
Q^2_{\rm min} = \frac{m_{e}^2x^2}{1-x},~~~Q^2_{\rm max} = (\theta_cE_{e})^2(1-x)+Q^2_{\rm min},\label{theta}
\end{equation}
where the electron scattering angle cut $\theta_c$ is determined by experiment \cite{wwa4, wwa5}.

The subtraction term  $f_{Q/P}(x_2, \mu_f)_{\rm SUB}$ in Eq.(\ref{eq1}) is defined as
\begin{equation}
f_{Q/p}(x_2, \mu_f)_{\rm SUB}= \int_{x_2}^1 f_{g/P}(x_2/y, \mu_f) f_{Q/g}(y, \mu_f) \frac{dy}{y},
\end{equation}
where $f_{Q/g}(y, \mu_f)$ is the $Q$-quark distribution function within an on-shell gluon and it can be expanded order by order in $\alpha_s$. At the $\alpha_s$-order, $f_{Q/g}(y, \mu_f)$ is given by
\begin{eqnarray}
f_{Q/g}(y,\mu_f)=\frac{\alpha_s(\mu_f)}{2\pi} {\rm{ln}} \frac{\mu_f^2}{m_Q^2}P_{g\to Q}(y),
\end{eqnarray}
where $P_{g\to Q}(y)=\frac{1}{2}(1-2y+2y^2)$ is the $g \to Q \bar{Q}$ splitting function.

The partonic hard cross section can be written as
\begin{equation}
d\hat{\sigma}(\gamma+i \to  (QQ')[n] + X)= \frac{{\bf \overline{\sum} }|\mathcal{M}|^2}{4\sqrt{(p_{1}+p_2)^2}|\vec{ p}_{1}|} d\Phi_j,
\end{equation}
where ${\bf \overline{\sum} }$ denotes the average of the spin and color states of initial particles and the sum of the color and spin states of all final particles, and $d\Phi_j$ represents the final $j$-body phase space element,
\begin{equation}
d\Phi_j = (2\pi)^4 \delta^4(p_1 + p_2 - \sum_{f=3}^{j+2} p_f) \prod_{f=3}^{j+2} \frac{d^3p_f}{(2\pi)^3 2p_f^0}
\end{equation}
and $\mathcal{M}$ is the total hard scattering amplitude
\begin{equation}
\mathcal{M}= { \sum_k } \mathcal{M}_k,
\end{equation}
where $k$ runs over the related Feynman diagrams.

\subsection{Feynman diagram and amplitude}

There are totally $24$ Feynman diagrams for the subprocess $\gamma + g \to  (QQ')[n] + \bar{Q} +\bar{Q'}$ ($k=24$) and $4$ Feynman diagrams for the subprocess $\gamma + Q \to (QQ')[n] + \bar{Q}$ ($k=4$). As for the subprocesses $\gamma + g \to (QQ)[n]  + \bar{Q}+\bar{Q}$ and $\gamma + Q \to (QQ)[n] + \bar{Q}$, there are another 24 and 4 diagrams coming from the exchanging of two identical quark lines inside the $(QQ)[n]$-quark pair. Practically, those diagrams are the same as the diagrams without exchanging, since we have set the relative velocity between the two $Q$ quarks to be zero, i.e., we have set $p_{31}=p_{32}=\frac{p_3}{2}$ for the production of $\Xi_{QQ}$ by applying the nonrelativistic approximation. There is a factor of $1/2!$ for the square of the amplitude due to the two identical quarks inside the $\langle QQ\rangle[n]$ diquark. Thus, we only need to calculate the $24$ and $4$ diagrams for $\gamma + g \to (QQ)[n]  + \bar{Q}+\bar{Q}$  and $\gamma + Q \to (QQ)[n]  + \bar{Q}$ subprocesses, and multiply a factor of $2^2/2!$ at the cross section level. Besides, there is another factor of $1/2$ for the $\gamma + g \to (QQ)[n]  + \bar{Q}+\bar{Q}$ subprocess coming from the two identical open antiquarks $\bar{Q}$ in the final 3-body phase space. The amplitudes for  $\gamma + g \to  (QQ')[n]  +  \bar{Q} +\bar{Q'}$ and  $\gamma + Q \to  (QQ')[n] + \bar{Q'}$ can be obtained directly from the Feynman diagrams. In order to describe the bound system of the doubly heavy baryon, we should apply the spin- and color-projection operators on the amplitude of $(QQ')[n]$-quark pair. For a detailed description on how to apply the projection operators on the amplitude of $(QQ')[n]$-quark pair and the calculation of the color factor of the production of heavy baryon, one can refer to Refs.\cite{lhecbc, xicc6}.

To implement the calculation, the FeynArts package~\cite{FA} is used to generate the Feynman diagrams and amplitudes; The FeynCalc~\cite{FC, FC1} and FeynCalcFormLink packages~\cite{FL} are used to handle the Dirac trace and $SU(N_c)$ algebraic calculations. Numerical integrations over 2- and 3-body phase spaces are performed by using the VEGAS~\cite{VEGAS} and FormCalc \cite{formcalc} packages.

\section{Numerical results and discussions} \label{sec3}

\subsection{Input parameters}

The matrix element $h_{\bar{\textbf{3}}}$ is related to the Schr$\ddot{o}$dinger wave function at the origin, $h_{\bar{\textbf{3}}}= |\Psi_{(QQ')}(0)|^2$~\cite{potential}. According to the velocity scaling rule of NRQCD~\cite{fpro3}, the color-sextuplet matrix element $h_{\textbf{6}}$ is at the same order of $h_{\bar{\textbf{3}}}$, and we take the usual choice of ${h_\textbf{6}} =h_{\bar{\textbf{3}}}$ to do our calculation. Since $h_{{\textbf{6}}/{\bar{\textbf{3}}}}$ is an overall factor, one can improve our results once more accurate $h_{{\textbf{6}}/{\bar{\textbf{3}}}}$ is known. The wave functions at the origin together with the heavy quark masses are taken as follows~\cite{xicc2, potential}:
\begin{eqnarray}
&& \quad \vert \Psi_{(cc)}(0) \vert^2=0.039 ~{\rm{GeV^3}},\vert \Psi_{(bc)}(0) \vert^2=0.065 ~{\rm{GeV^3}},\nonumber \\
&&\vert \Psi_{(bb)}(0) \vert^2=0.152 ~{\rm{GeV^3}},
m_b=5.1~{\rm{GeV}}, m_c=1.8 ~{\rm{GeV}}.\nonumber
\end{eqnarray}
The electron mass $m_e=0.51\times 10^{-3}$ GeV is adopted and the fine-structure constant is fixed as $\alpha = 1/137$. The electron scattering angle cut $\theta_c$ is chosen as 32~{\rm{mrad}} which is consistent with the choices of Refs.\cite{theta1, theta2}. The renormalization and factorization scales are set to be the transverse mass of $\Xi_{QQ'}$, $\mu_r = \mu_f = M_T$, where $M_T = \sqrt{p_T^2+M^2}$ with $M$ being the mass of $\Xi_{QQ'}$. Here $M=m_Q+m_{Q'}$, which ensures the gauge invariance of the hard scattering amplitude. The PDF of the incident quark in hadron is taken as CT10NLO~\cite{CT10NLO}, and correspondingly, the next-to-leading order $\alpha_s$-running with $\Lambda_{QCD}^{(4)}=326~ {\rm MeV}$ ($\Lambda_{QCD}^{(5)}=226~ {\rm MeV}$) is adopted.

\subsection{Basic results}

To shorten the notation, we denote the cross sections for the production of $\langle QQ'\rangle[^1S_0]_{{\bar{\textbf{3}}}/{\textbf{6}}}$ and $\langle QQ'\rangle[^3S_1]_{{\bar{\textbf{3}}}/{\textbf{6}}}$ via the $\gamma+i$ channels as $\sigma_{\gamma i}$ and $\sigma_{\gamma i}^*$, where $i=g, c, b$ respectively. We use $\langle bc\rangle_{\bar{\textbf{3}}}$ and ${\langle bc\rangle_{\textbf{6}}}$ to represent the production of $\langle bc\rangle[^1S_0/^3S_1]_{\bar{\textbf{3}}}$ and $\langle bc\rangle[^1S_0/^3S_1]_{\textbf{6}}$. We use $\langle QQ \rangle_{\textbf{c}}$ to represent the production of $\langle QQ \rangle[^3S_1]_{\bar{\textbf{3}}}$ or $\langle QQ \rangle[^1S_0]_{\textbf{6}}$, since $\langle QQ \rangle[^3S_1]_{\textbf{6}}$ and $\langle QQ \rangle[^1S_0]_{\bar{\textbf{3}}}$ are forbidden.

\begin{widetext}
\begin{center}
\begin{table}[htb]
\caption{Total cross sections (in unit pb) for the photoproduction of $\Xi_{QQ'}$ at the LHeC and FCC-$ep$ colliders.}
\begin{tabular}{c| c c c c| c c c c| c c c c|cccc}
\hline
- & \multicolumn{4}{c|}{$\sqrt{S}=1.30$ TeV} & \multicolumn{4}{|c|}{$\sqrt{S}=1.98$ TeV}& \multicolumn{4}{|c}{$\sqrt{S}=7.07$ TeV}& \multicolumn{4}{|c}{$\sqrt{S}=10.00$ TeV}\\
\hline - & $ \langle bb \rangle_{\textbf{c}}$ & $ \langle cc \rangle_{\textbf{c}}$ & $ \langle bc \rangle_{\bar{\textbf{3}}}$ & $ \langle bc \rangle_{{\textbf{6}}} $& $ \langle bb \rangle_{\textbf{c}}$ & $ \langle cc \rangle_{\textbf{c}}$ & $ \langle bc \rangle_{\bar{\textbf{3}}}$ & $ \langle bc \rangle_{{\textbf{6}}} $& $ \langle bb \rangle_{\textbf{c}}$ & $ \langle cc \rangle_{\textbf{c}}$ & $ \langle bc \rangle_{\bar{\textbf{3}}}$ & $ \langle bc \rangle_{{\textbf{6}}} $&$ \langle bb \rangle_{\textbf{c}}$ & $ \langle cc \rangle_{\textbf{c}}$ & $ \langle bc \rangle_{\bar{\textbf{3}}}$ & $ \langle bc \rangle_{{\textbf{6}}} $ \\
 $\sigma_{\gamma g}$  &  $0.02$  & $21.61$   &  $1.55$    &$1.06$  &
                         $0.04$  & $32.81$   &  $2.58$    &$1.76$  &
                         $0.16$  & $82.81$   &  $7.99$    &$5.39$  &
                         $0.22$  & $108.43$   &  $10.95$    &$7.37$  \\
 $\sigma_{\gamma c}$  &  -     & $42.91$   &  $0.27$    &$0.13$&
                         -     & $62.28$   &  $0.41$    &$0.20$&
                         -     & $140.60$  &  $1.00$    &$0.50$  &
                         -     & $178.84$   &  $1.29$    &$0.65$  \\
 $\sigma_{\gamma b}$  &  $0.02$  & -       &  $3.84$    &$1.92$&
                         $0.04$  & -       &  $6.38$    &$3.19$&
                         $0.15$  & -       &  $20.01$   &$10.01$ &
                         $0.21$  & -       &  $27.59$    &$13.80$  \\
 $\sigma_{\gamma g}^*$&  $0.29$  & $241.67$  &  $5.09$    &$4.32$&
                         $0.50$  & $360.75$  &  $8.43$    &$7.12$&
                         $1.75$  & $877.57$  &  $25.73$   &$21.53$  &
                         $2.46$  & $1139.59$  &  $35.17$    &$29.36$  \\
 $\sigma_{\gamma c}^*$&  -     & $574.42$  &  $1.26$    &$0.63$&
                         -     & $829.44$  &  $1.91$    &$0.95$&
                         -     & $1854.55$ &  $4.62$    &$2.31$&
                         -     & $2354.47$  &  $5.95$    &$2.97$  \\
 $\sigma_{\gamma b}^*$&  $0.34$  & -       &  $17.48$   &$8.74$&
                         $0.59$  & -       &  $28.98$   &$14.49$&
                         $2.04$  & -       &  $90.71$   &$45.35$  &
                         $2.86$  & -       &  $125.02$   &$62.51$  \\
\hline        Total         &0.67&880.61&\multicolumn{2}{c|}{46.29}&1.17&1285.28 &\multicolumn{2}{c|}{76.40}&4.10&2955.53&\multicolumn{2}{c|}{235.15} &5.75&3781.33&\multicolumn{2}{c}{322.63}  \\
\hline
\end{tabular}
\label{color3total}
\end{table}
\end{center}
\end{widetext}

We present the cross sections for the photoproduction of $\Xi_{QQ'}$ at the electron-hadron colliders under various production channels in Table~\ref{color3total}. Here, to show how the cross section depends on the electron-hadron colliding energy, we present the numerical results at two $ep$ colliders with four colliding energies, i.e., $\sqrt{S}=1.30~{\rm TeV}$ or $1.98~{\rm TeV}$ for LHeC which corresponds to $E_e=60$ or $140~ {\rm GeV}$ and $E_P=7~ {\rm TeV}$~\cite{LHeC}, and $\sqrt{S}=7.07~{\rm TeV}$ or $10.00~{\rm TeV}$ for the future circular collider based $ep$ collider (FCC-$ep$) which corresponds to $E_e=250$ or $500~ {\rm GeV}$ and $E_P=50~ {\rm TeV}$~\cite{fcc}.

\begin{figure}[htb]
\includegraphics[width=0.45\textwidth]{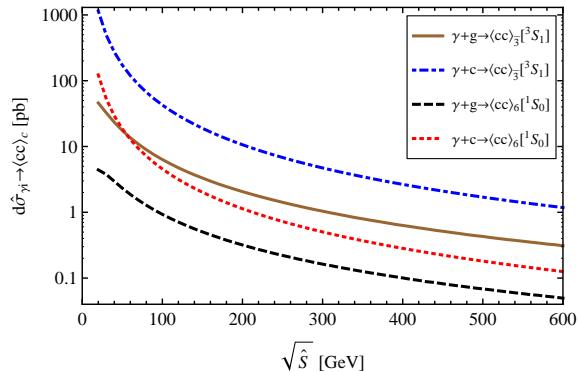}
\caption{Partonic cross sections for the photoproduction of $\Xi_{QQ'}$ versus $\sqrt{ \hat{s}}$.} \label{subdist}
\end{figure}

Table~\ref{color3total} shows
\begin{itemize}
\item For the same production channel, the $[^3S_1]_{\bar{\textbf{3}}}$ diquark state provides the largest contribution for the production cross section of $\Xi_{QQ'}$. For the $\Xi_{cc}$ and $\Xi_{bb}$ production, the color-sextuplet $[^1S_0]_{\textbf{6}}$ diquark state gives about $7\%-8\%$ contribution to the total cross section of the same production channel. For the $\Xi_{bc}$ production, contributions from other diquark states are also sizable, especially, the cross section of $[^3S_1]_{\textbf{6}}$ diquark state is close to that of $[^3S_1]_{\bar{\textbf{3}}}$. Thus a careful discussion of all diquark configurations are helpful for a sound prediction of the doubly heavy baryon production.

\item In addition to the usually considered $\gamma+g$ channel, the extrinsic heavy quark mechanism via the $\gamma+Q$ channel shall also provide sizable contribution to the production cross section. For example, the $\gamma+c$ channel is even dominant over the $\gamma+g$ channel for the $\Xi_{cc}$ production. This dominance, as shall be shown later, is caused by the dominance of the $\gamma+Q$ channel in low $p_T$ region.

\item By summing up all the mentioned diquark configurations and production channels, we obtain $\sigma_{\langle cc  \rangle}^{\rm{Total}}\gg\sigma_{\langle bc \rangle}^{\rm{Total}}\gg\sigma_{\langle bb \rangle}^{\rm{Total}}$, which agrees with the observation of the doubly heavy baryon production at the ILC~\cite{ggxiqq}.

\item The total cross section increases with the increment of electron-proton colliding energy, differing to the doubly heavy baryon production at the ILC~\cite{ggxiqq}, where the total cross section decreases with the increment of photon-photon colliding energy. However, it is found that at the partonic level, the production cross sections for the $\gamma+\gamma \to \Xi_{QQ'}+X$ and $\gamma+i \to \Xi_{QQ'}+X$ behave closely, both of which shall decrease with the increment of the subprocess colliding energy.
    For example, we present the dependence of the cross sections for the subprocess $\gamma+i \to \Xi_{cc}+X$ in Fig.\ref{subdist}.

\end{itemize}

We have found that the photoproduction of $\Xi_{QQ'}$ are similar under various electron-proton collision energies. In the following, we take $\sqrt{S}=1.30~ {\rm TeV}$ as an explicit example to show the photoproduction of $\Xi_{QQ'}$ in detail.

\begin{figure*}[htb]
\includegraphics[width=0.4\textwidth]{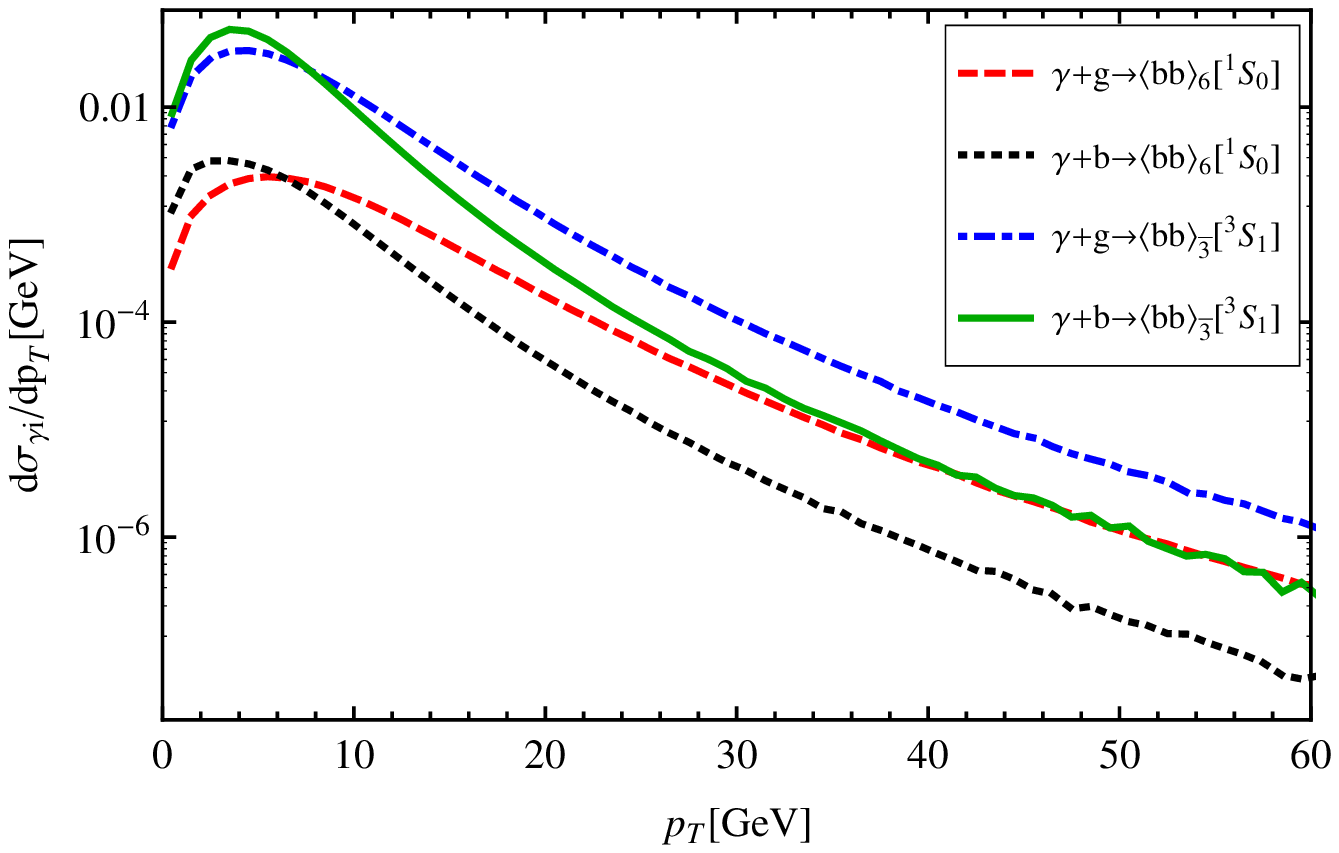}
\includegraphics[width=0.4\textwidth]{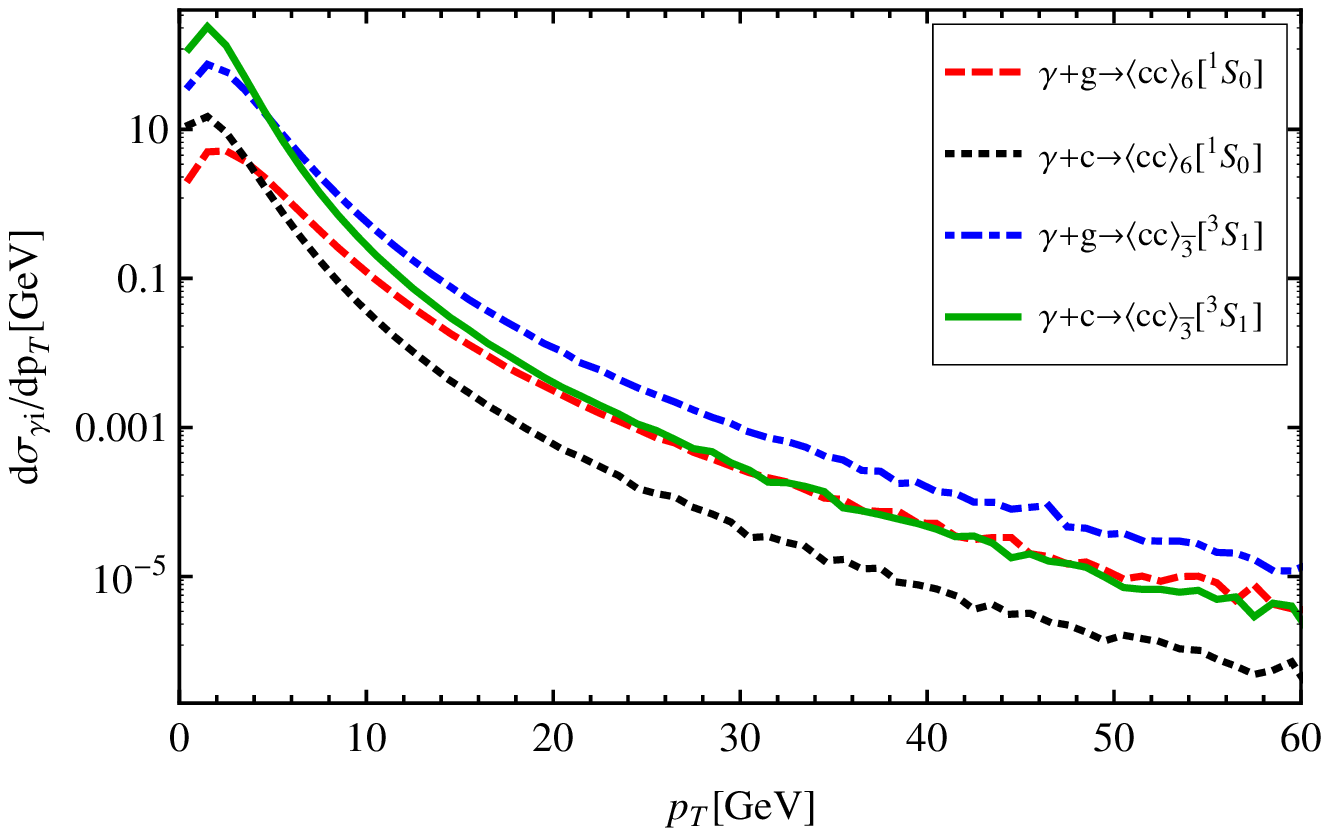}
\includegraphics[width=0.4\textwidth]{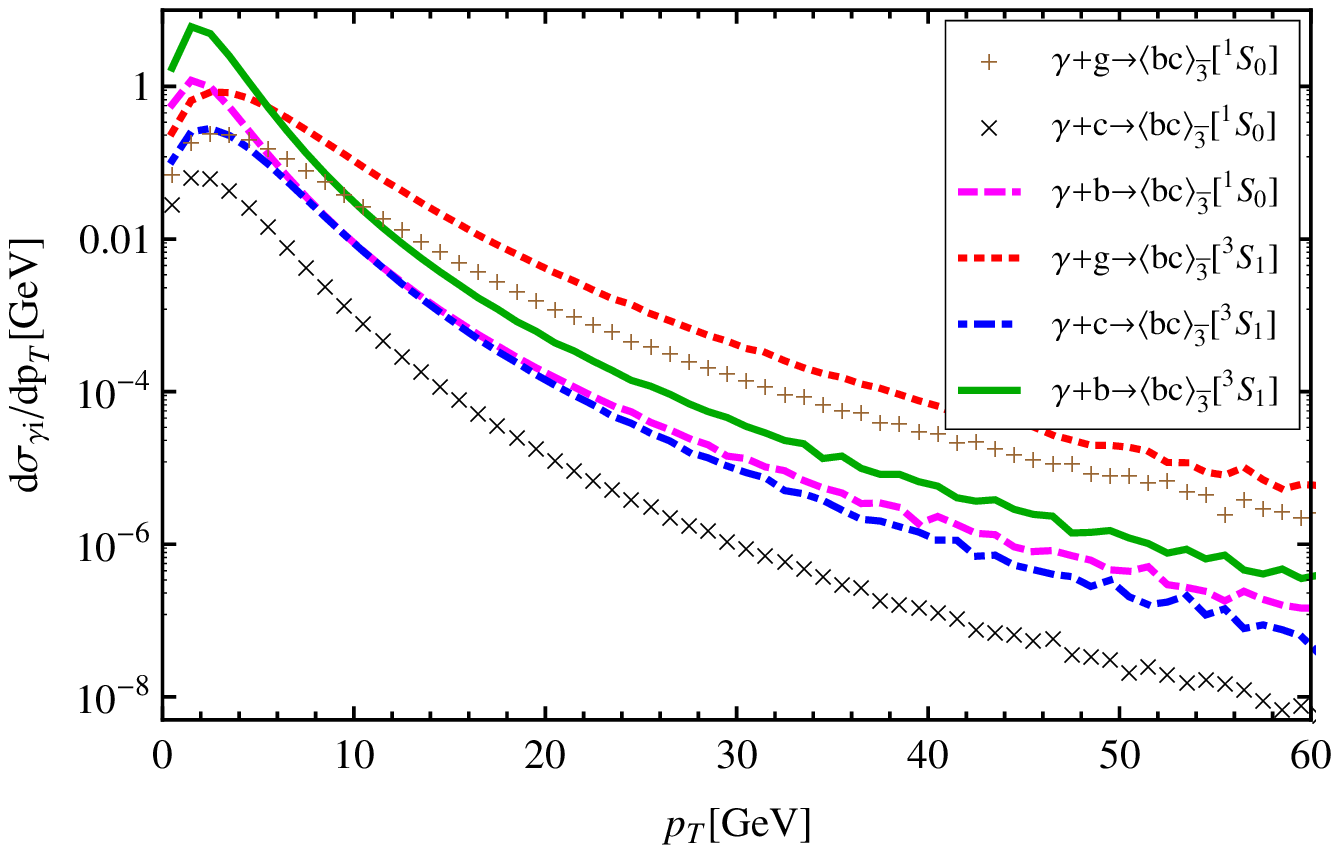}
\includegraphics[width=0.4\textwidth]{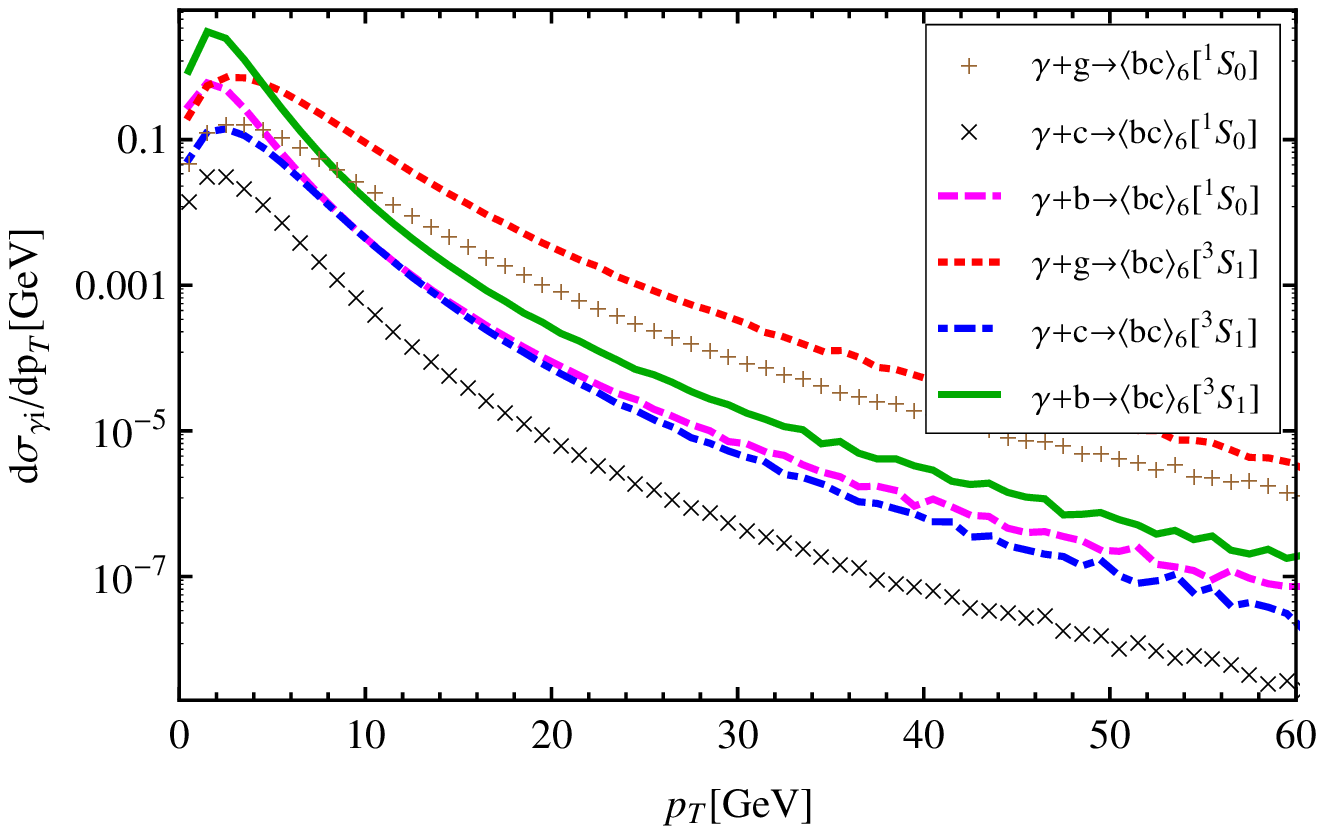}
\caption{Transverse momentum distributions for the photoproduction of $\Xi_{QQ'}$ at the $\sqrt{S}=1.30~{\rm TeV}$ LHeC.}
\label{figspt}
\end{figure*}

We present the transverse momentum ($p_T$) distributions for the photoproduction of ${\Xi_{QQ'}}$ in Fig.\ref{figspt}. The $p_T$ distribution for each channel has a peak for $p_T \sim {\cal O}(1)~{\rm GeV}$ and then drops down logarithmically. The $p_T$ distributions of the $\gamma+ c$ and $\gamma+\bar{b}$ channels descend more quickly than those of $\gamma+ g$ channels in high $p_T$ region. For the ${\Xi_{QQ'}}$ production via the same diquark configuration, the $\gamma+Q$ channel dominates the $\gamma+g$ channel in small $p_T$ region, which explains sizable total cross section of $\gamma+Q$ channel as shown in Table~\ref{color3total}. We also observe that the $p_T$ distributions of $\Xi_{bb}$ decrease more slowly than those of $\Xi_{bc}$ and $\Xi_{cc}$ with the increment of $p_T$, and thus the $\Xi_{bb}$ events may be comparable to $\Xi_{cc}$ and $\Xi_{bc}$ at high $p_T$ region, although the total cross section of $\Xi_{bb}$ is much smaller than those of $\Xi_{cc}$ and $\Xi_{bc}$.

\begin{figure*}[htb]
\includegraphics[width=0.4\textwidth]{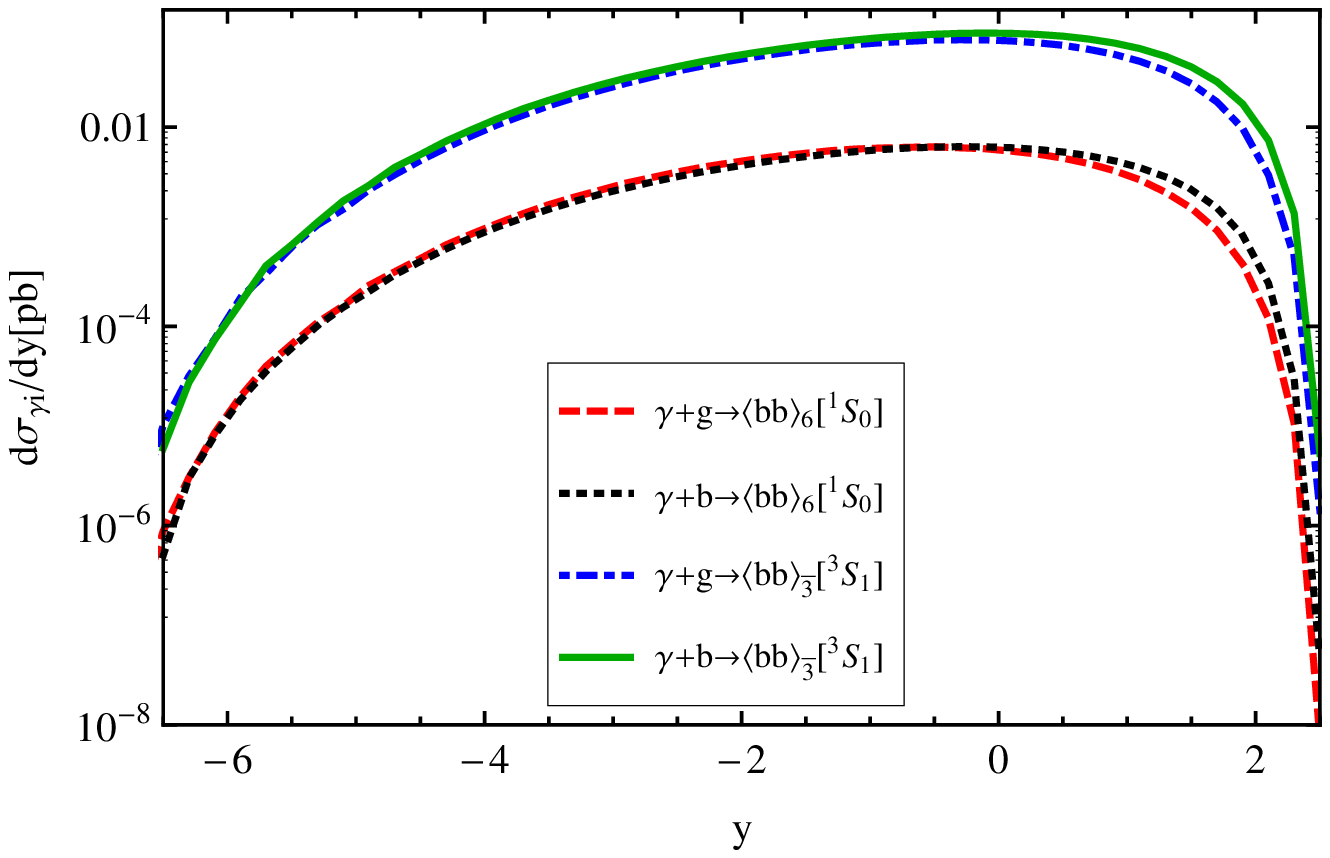}
\includegraphics[width=0.4\textwidth]{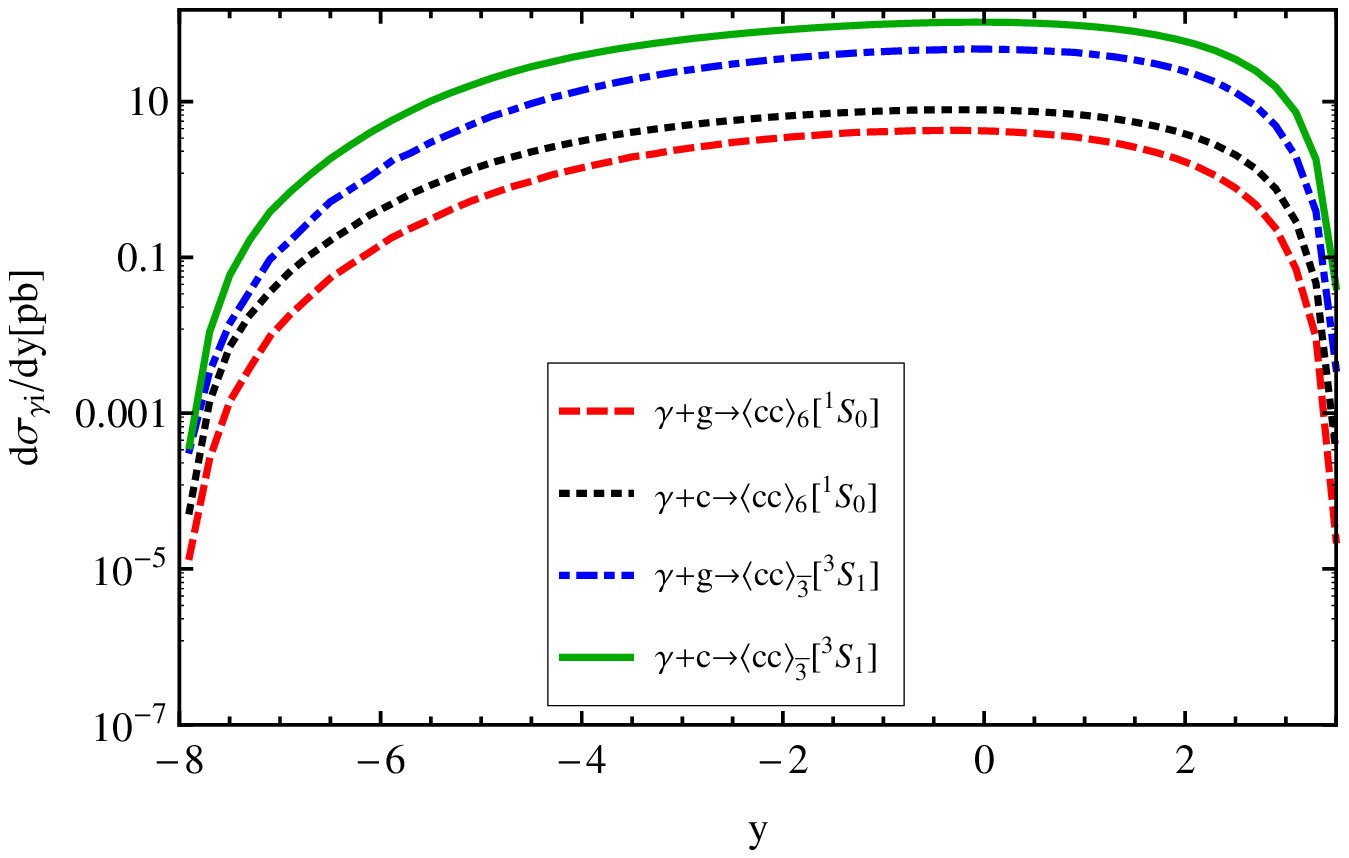}
\includegraphics[width=0.4\textwidth]{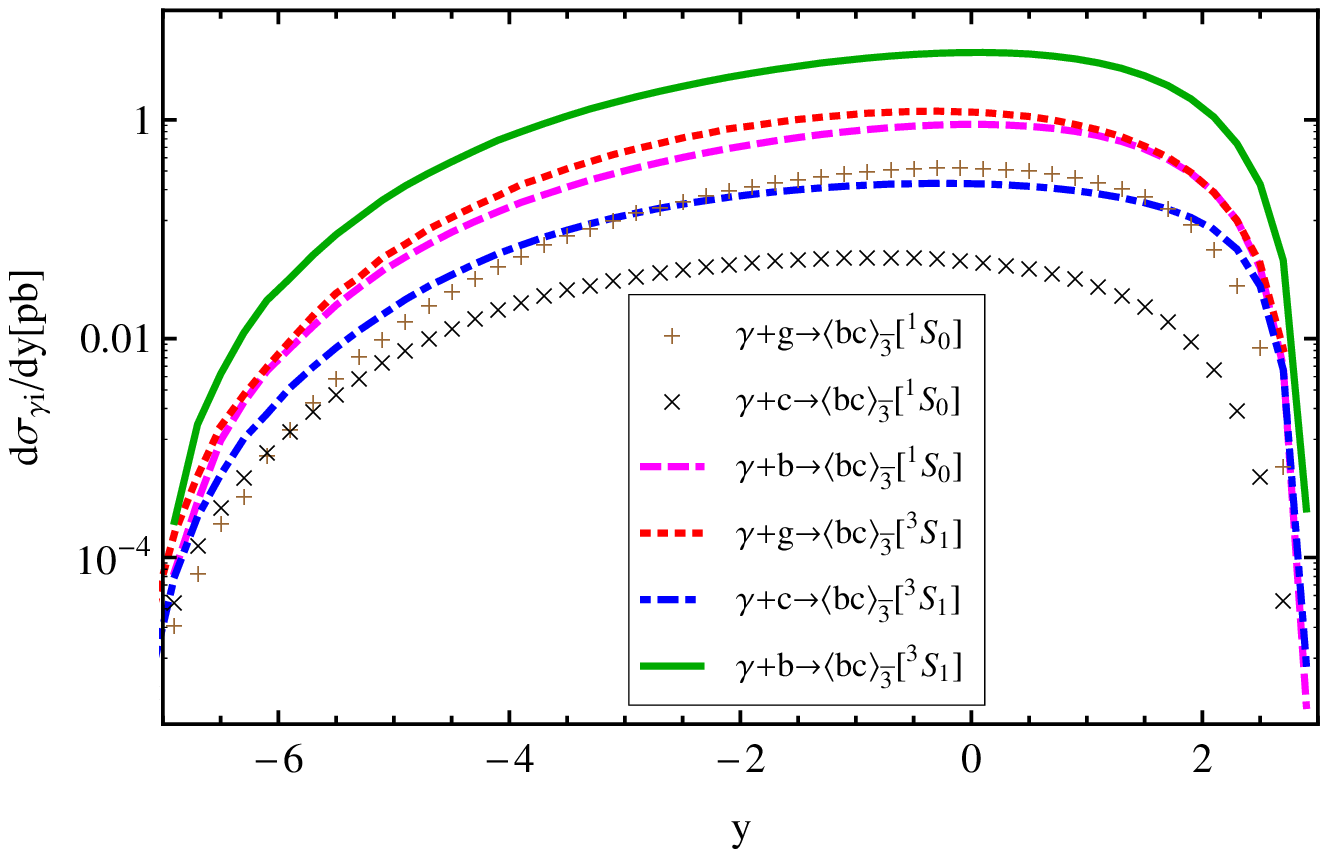}
\includegraphics[width=0.4\textwidth]{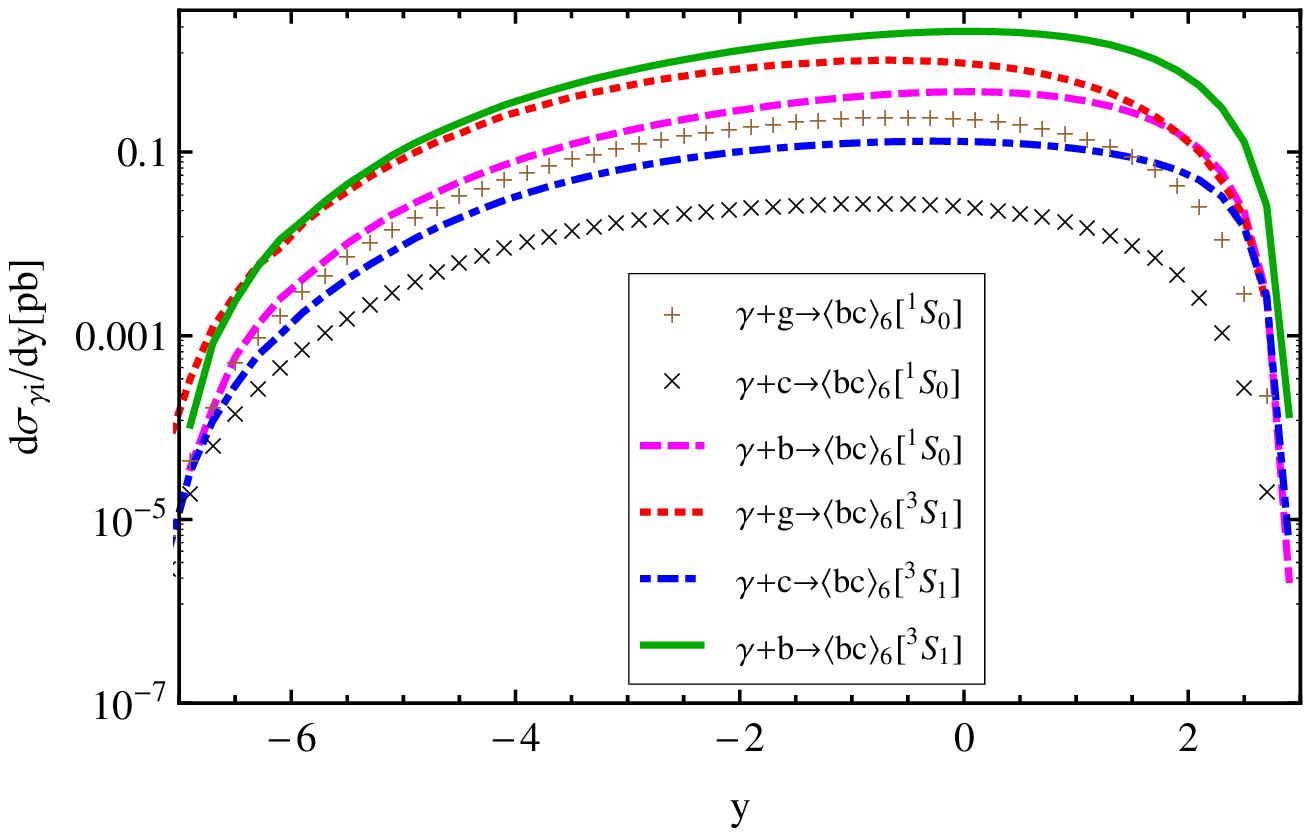}
\caption{Rapidity distributions for the photoproduction of $\Xi_{QQ'}$  at the $\sqrt{S}=1.30~{\rm TeV}$ LHeC.} \label{figsy}
\end{figure*}

We present the rapidity ($y$) distributions for the photoproduction of ${\Xi_{QQ'}}$ in Fig.\ref{figsy}. The asymmetry of the rapidity distributions of $\Xi_{QQ'}$ clearly shows that the dominant contribution appears in the region of $y<0$. The $z$-axis is defined in the direction of the electron beam, thus the fact of $y<0$ implies the parton $i$ from the proton is more energetic than the photon, and most of $\Xi_{QQ'}$ events tend to be produced in the direction of the proton beam.

\begin{widetext}
\begin{center}
\begin{table}[htb]
\caption{Total cross sections (in unit pb) for the photoproduction of $\Xi_{QQ'}$  at the $\sqrt{S}=1.30$ TeV LHeC under various $p_T$ cuts.}
\begin{tabular}{c| c c c c| c c c c| c c c c}
\hline
- & \multicolumn{4}{c|}{$p_T\geq1$ GeV} & \multicolumn{4}{|c|}{$p_T\geq3$ GeV}& \multicolumn{4}{|c}{$p_T\geq5$ GeV}\\
\hline - & $ \langle bb \rangle_{\textbf{c}}$ & $ \langle cc \rangle_{\textbf{c}}$ & $ \langle bc \rangle_{\bar{\textbf{3}}}$ & $ \langle bc \rangle_{{\textbf{6}}} $& $ \langle bb \rangle_{\textbf{c}}$ & $ \langle cc \rangle_{\textbf{c}}$ & $ \langle bc \rangle_{\bar{\textbf{3}}}$ & $ \langle bc \rangle_{{\textbf{6}}} $& $ \langle bb \rangle_{\textbf{c}}$ & $ \langle cc \rangle_{\textbf{c}}$ & $ \langle bc \rangle_{\bar{\textbf{3}}}$ & $ \langle bc \rangle_{{\textbf{6}}} $ \\
 $\sigma_{\gamma g}$  &  $0.02$  & $19.46$  &  $1.47$    &  $1.01$  &
                         $0.02$  & $9.34$   &  $1.03$    &  $0.71$  &
                         $0.02$  & $3.26$   &  $0.57$     & $0.39$  \\
 $\sigma_{\gamma c}$  &  -     & $31.45$  &  $0.24$    & $0.12$&
                         -     & $7.48$   &  $0.11$    &$0.05$&
                         -     & $1.52$   &  $0.03$    &$0.02$  \\
 $\sigma_{\gamma b}$  &$0.02$&-&$3.27$&$1.63$&
                       $0.02$&-&$1.08$&$0.54$&
                       $0.01$&-&$0.28$&$0.14$  \\
 $\sigma_{\gamma g}^*$&$0.28$&$203.55$&$5.09$&$4.11$&
                       $0.23$&$70.07$&$3.35$&$2.85$&
                       $0.16$&$18.81$&$1.82$&$1.55$  \\
 $\sigma_{\gamma c}^*$&-&$454.38$&$1.16$&$0.58$&
                       -&$82.11$&$0.62$&$0.31$&
                       -&$13.30$&$0.24$&$0.12$  \\
 $\sigma_{\gamma b}^*$&$0.33$&-&$15.73$&$7.87$&
                       $0.26$&-&$4.76$&$2.38$&
                       $0.15$&-&$1.11$&$0.56$  \\
\hline        Total         &$0.65$&$708.84$&\multicolumn{2}{c|}{42.28}&$0.53$&$169.00$&\multicolumn{2}{c|}{17.79}&$0.34$&$36.89$&\multicolumn{2}{c}{6.83}  \\
\hline
\end{tabular}
\label{tabsecvspt}
\end{table}
\end{center}
\end{widetext}

\begin{widetext}
\begin{center}
\begin{table}[htb]
\caption{Total cross sections (in unit pb) for the photoproduction of $\Xi_{QQ'}$  at the $\sqrt{S}=1.30$ TeV LHeC under various $y$ cuts.}
\begin{tabular}{c| c c c c| c c c c| c c c c}
\hline
- & \multicolumn{4}{c|}{$|y|\leq1$} & \multicolumn{4}{|c|}{$|y|\leq2$}& \multicolumn{4}{|c}{$|y|\leq3$}\\
\hline - & $ \langle bb \rangle_{\textbf{c}}$ & $ \langle cc \rangle_{\textbf{c}}$ & $ \langle bc \rangle_{\bar{\textbf{3}}}$ & $ \langle bc \rangle_{{\textbf{6}}} $& $ \langle bb \rangle_{\textbf{c}}$ & $ \langle cc \rangle_{\textbf{c}}$ & $ \langle bc \rangle_{\bar{\textbf{3}}}$ & $ \langle bc \rangle_{{\textbf{6}}} $& $ \langle bb \rangle_{\textbf{c}}$ & $ \langle cc \rangle_{\textbf{c}}$ & $ \langle bc \rangle_{\bar{\textbf{3}}}$ & $ \langle bc \rangle_{{\textbf{6}}} $ \\
 $\sigma_{\gamma g}$  &  $0.01$  & $8.11$  &  $0.71$    &  $0.45$  &
                         $0.02$  & $14.52$   &  $1.21$    &  $0.77$  &
                         $0.02$  & $18.30$   &  $1.42$     & $0.93$  \\
 $\sigma_{\gamma c}$  &  -     & $15.18$  &  $0.10$    & $0.05$&
                         -     & $27.65$   &  $0.18$    &$0.09$&
                         -     & $35.46$   &  $0.22$    &$0.11$  \\
 $\sigma_{\gamma b}$  &$0.01$&-&$1.73$&$0.86$&
                       $0.02$&-&$2.94$&$1.47$&
                       $0.02$&-&$3.48$&$1.74$  \\
 $\sigma_{\gamma g}^*$&$0.14$&$91.18$&$2.23$&$1.76$&
                       $0.23$&$165.01$&$3.80$&$3.01$&
                       $0.26$&$208.64$&$4.56$&$3.72$  \\
 $\sigma_{\gamma c}^*$&-&$203.34$&$0.50$&$0.25$&
                       -&$373.08$&$0.90$&$0.45$&
                       -&$480.93$&$1.11$&$0.56$  \\
 $\sigma_{\gamma b}^*$&$0.16$&-&$7.85$&$3.93$&
                       $0.27$&-&$13.42$&$6.71$&
                       $0.31$&-&$15.84$&$7.92$  \\
\hline        Total         &$0.32$&$317.81$&\multicolumn{2}{c|}{20.42}&$0.54$&$480.26$&\multicolumn{2}{c|}{34.95}&$0.61$&$743.33$&\multicolumn{2}{c}{41.61}  \\
\hline
\end{tabular}
\label{tabsecvsy}
\end{table}
\end{center}
\end{widetext}

We present the cross sections under various $p_T$ and $y$ cuts in Tables~\ref{tabsecvspt} and~\ref{tabsecvsy}. Table~\ref{tabsecvspt} shows the cross section of $\Xi_{bc}$ and $\Xi_{cc}$ under various of $p_T$ cuts are more sensitive than that of $\Xi_{bb}$, which is consistent with Fig.~\ref{figspt}. Table~\ref{tabsecvspt} shows notable $\Xi_{bc}$ and $\Xi_{cc}$ production rate can be generated even with a large $p_T$ cut.

\begin{figure*}[htb]
\includegraphics[width=0.4\textwidth]{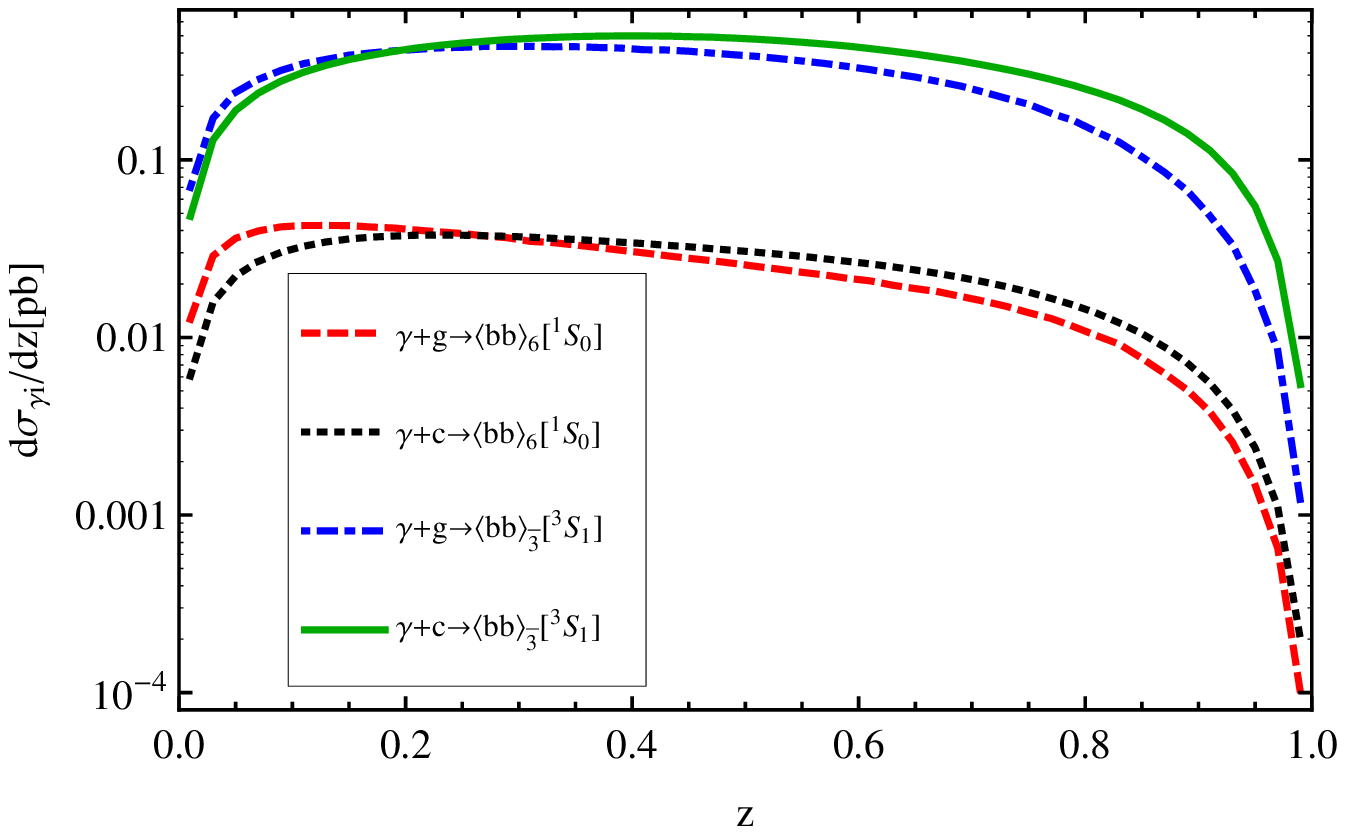}
\includegraphics[width=0.4\textwidth]{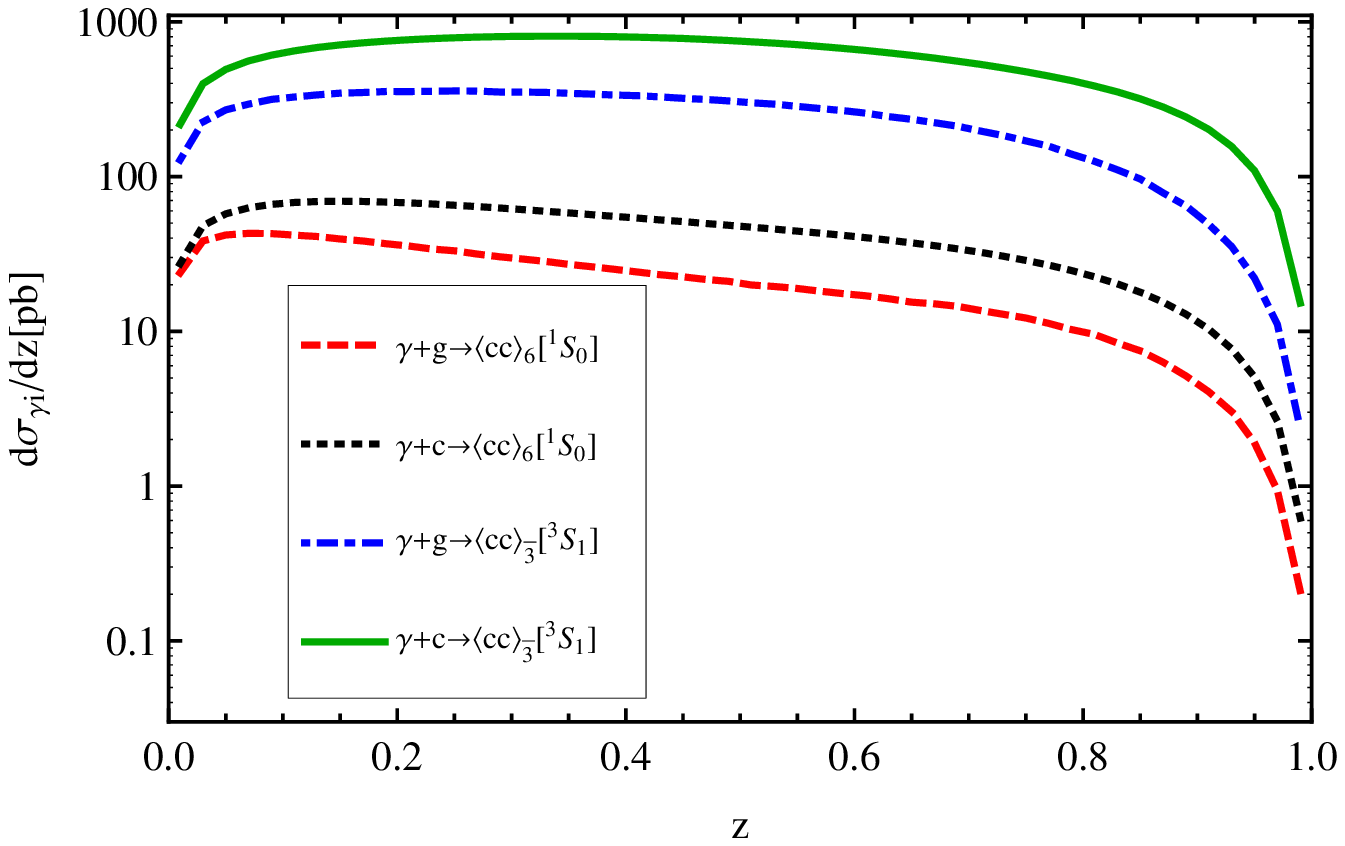}
\includegraphics[width=0.4\textwidth]{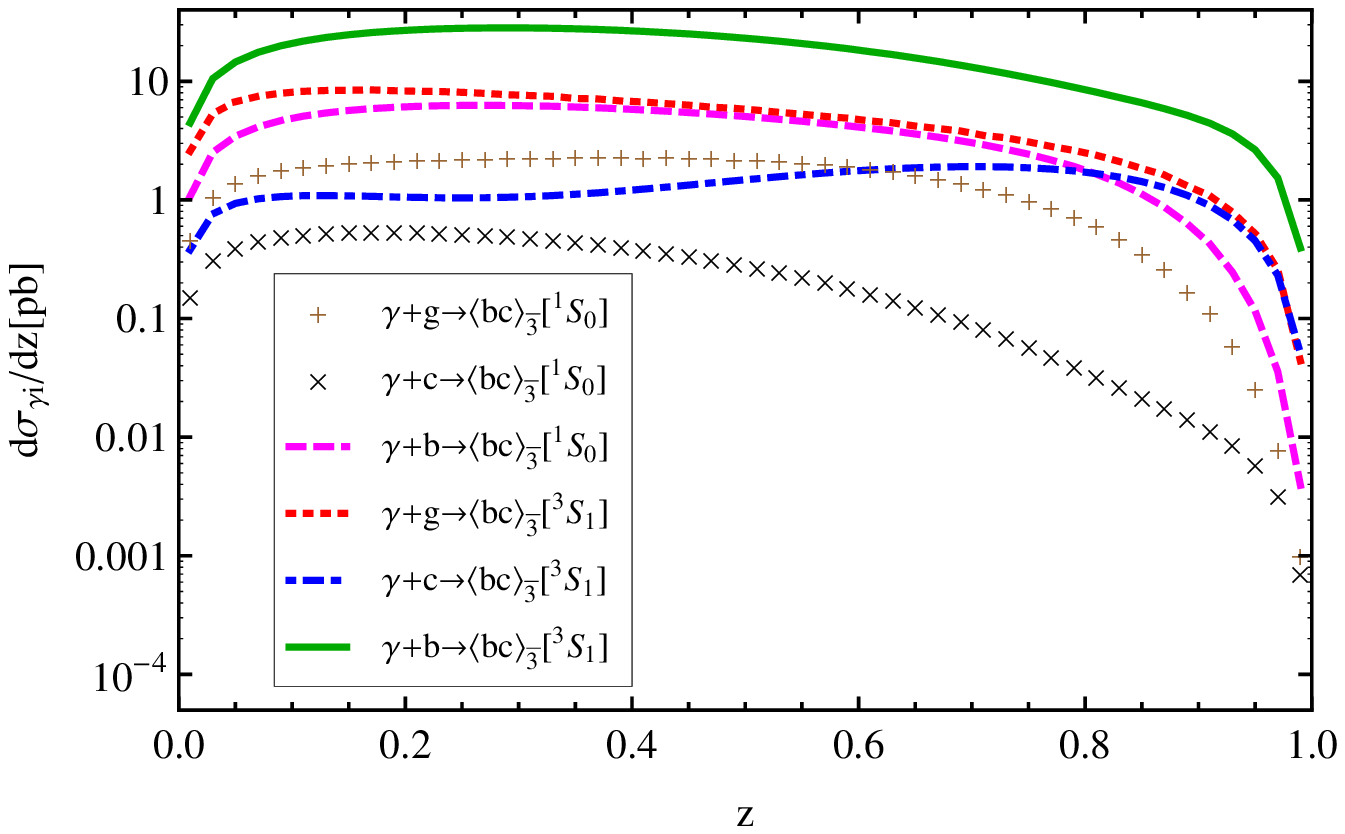}
\includegraphics[width=0.4\textwidth]{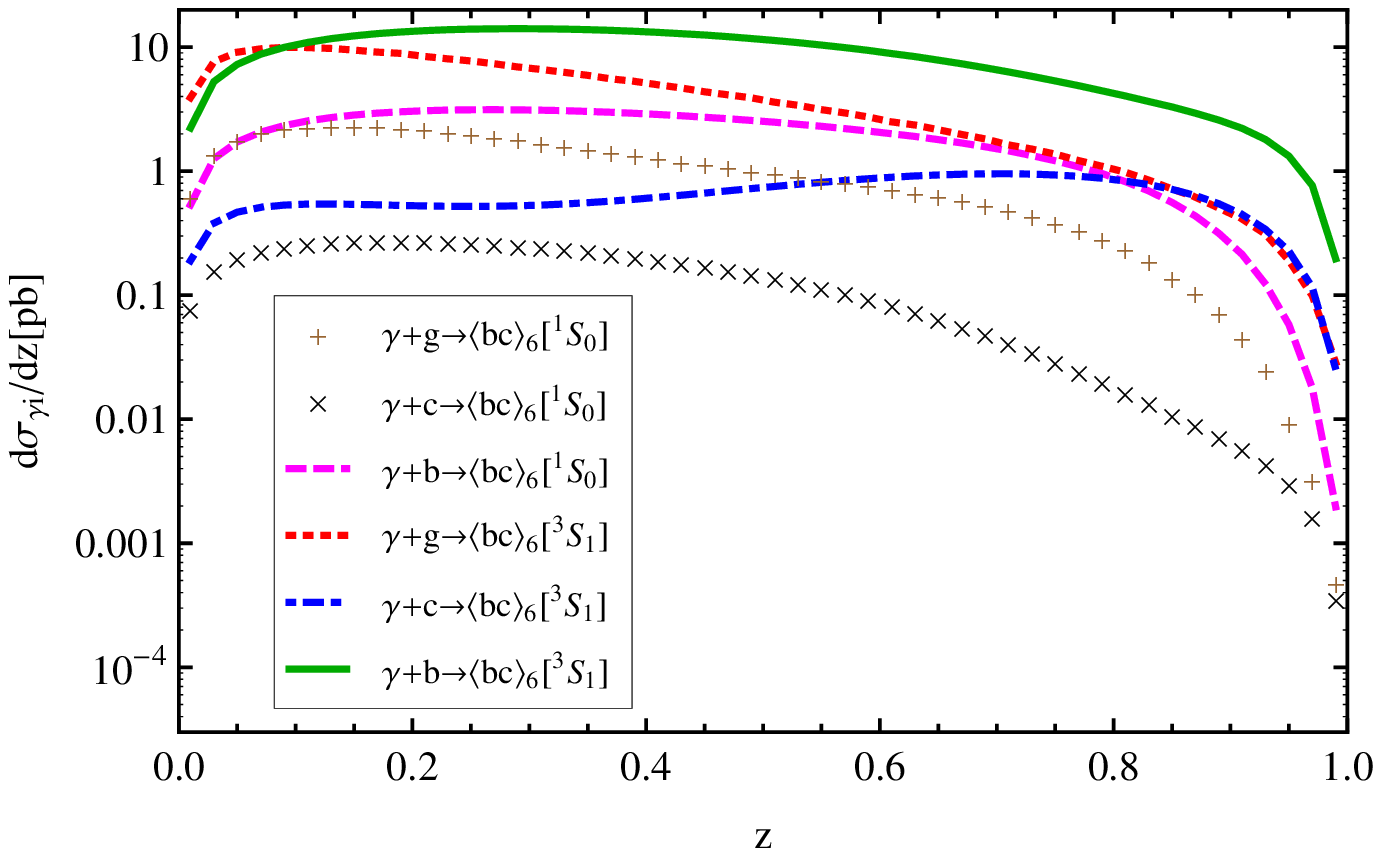}
\caption{$z$ distributions for the photoproduction of $\Xi_{QQ'}$ at the $\sqrt{S}=1.30$ TeV LHeC.} \label{figsz}
\end{figure*}

\begin{table}
\caption{Variation for the total cross sections (in unit pb) for the photoproduction of $\Xi_{QQ'}$ at the $\sqrt{S}=1.30$ TeV LHeC under the cut $0.3\lesssim z \lesssim0.9$.}
\begin{center}
\begin{tabular}{c| c c c c}
\hline - & \multicolumn{4}{c}{$0.3\lesssim z \lesssim0.9$} \\
\hline - & $ \langle bb \rangle_{\textbf{c}}$ & $ \langle cc \rangle_{\textbf{c}}$ & $ \langle bc \rangle_{\bar{\textbf{3}}}$ & $ \langle bc \rangle_{{\textbf{6}}} $\\
 $\sigma_{\gamma g}$     &  $0.01$  & $10.36$  &  $0.98$    &  $0.47$    \\
 $\sigma_{\gamma c}$     &  -     & $23.75$  &  $0.12$    & $0.06$  \\
 $\sigma_{\gamma b}$     &  $0.01$  &-       &$2.32$      &$1.16$  \\
 $\sigma_{\gamma g}^*$   &$0.18$&$144.82$&$0.98$&$1.78$ \\
 $\sigma_{\gamma c}^*$   &-&$368.98$&$0.92$&$0.46$  \\
 $\sigma_{\gamma b}^*$   &$0.23$&-&$10.65$&$5.33$  \\
\hline        Total         &0.43&547.91& \multicolumn{2}{c}{25.23} \\
\hline
\end{tabular}
\label{zcut}
\end{center}
\end{table}

Finally, we draw the differential cross sections $d\sigma/dz$ for various diquark configurations and production channels in Fig.\ref{figsz}, where $z=\frac{p_{\Xi}\cdot p_P}{p_{\gamma} \cdot p_P}$. Here, $p_{\gamma}$, $p_P$, and $p_{\Xi}$ are four momenta of the photon, proton, and $\Xi_{QQ'}$ respectively. For elastic/diffractive events, $z$ is close to $1$~\cite{jpsinlo}, besides, at low $z$ region, the resolved effect should be taken into consideration~\cite{H1}. For the prediction of inelastic direct photoproduction, a proper $z$ cut should be taken. We take $0.3\lesssim z \lesssim0.9$, which accounts for a clean sample of inelastic direct photoproduction~\cite{jpsinlo,H1}. The cross sections for various diquark configurations and production channels under this cut are listed in Table~\ref{zcut}.

\subsection{Theoretical uncertainties}

In this subsection, we make a discussion on the uncertainties caused by the $c$-quark mass, the $b$-quark mass, the renormalization (factorization) scale and the electron scattering angle cut $\theta_c$. All the other parameters are fixed as their center values when discussing the uncertainty from in variation of one parameter.

\begin{widetext}
\begin{center}
\begin{table}[htb]
\caption{Variations for the total cross sections  (in units pb) for the photoproduction of $\Xi_{QQ'}$ at the $\sqrt{S}=1.30$ TeV LHeC with $m_c = 1.80 \pm 0.10 \, \rm{GeV}$ and $m_b = 5.10 \pm 0.20 \, \rm{GeV}$. $m_b$ is fixed to 5.10 GeV when discussing the uncertainty from $m_c$, and $m_c$ is fixed to 1.8 GeV when discussing the uncertainty from $m_b$.}
\begin{tabular}{c| c c c c| c c c c| c c c c|cccc}
\hline
- & \multicolumn{4}{c|}{$m_c=1.7$ GeV} & \multicolumn{4}{|c|}{$m_c=1.9$ GeV}& \multicolumn{4}{|c}{$m_b=4.9$ GeV}& \multicolumn{4}{|c}{$m_b=5.3$ GeV}\\
\hline - & $ \langle bb \rangle_{\textbf{c}}$ & $ \langle cc \rangle_{\textbf{c}}$ & $ \langle bc \rangle_{\bar{\textbf{3}}}$ & $ \langle bc \rangle_{{\textbf{6}}} $& $ \langle bb \rangle_{\textbf{c}}$ & $ \langle cc \rangle_{\textbf{c}}$ & $ \langle bc \rangle_{\bar{\textbf{3}}}$ & $ \langle bc \rangle_{{\textbf{6}}} $& $ \langle bb \rangle_{\textbf{c}}$ & $ \langle cc \rangle_{\textbf{c}}$ & $ \langle bc \rangle_{\bar{\textbf{3}}}$ & $ \langle bc \rangle_{{\textbf{6}}} $&$ \langle bb \rangle_{\textbf{c}}$ & $ \langle cc \rangle_{\textbf{c}}$ & $ \langle bc \rangle_{\bar{\textbf{3}}}$ & $ \langle bc \rangle_{{\textbf{6}}} $ \\
 $\sigma_{\gamma g}$  &  $0.02$  & $30.90$   &  $1.84$    &$1.28$  &
                         $0.02$  & $15.38$   &  $1.31$    &$0.89$  &
                         $0.03$  & $21.61$   &  $1.78$    &$1.21$  &
                         $0.02$  & $21.61$   &  $1.35$    &$0.94$  \\
 $\sigma_{\gamma c}$  &  -     & $56.59$   &  $0.28$    &$0.14$  &
                         -     & $32.84$   &  $0.25$    &$0.13$  &
                         -     & $42.91$   &  $0.32$    &$0.15$  &
                         -     & $42.91$   &  $0.23$    &$0.11$  \\
 $\sigma_{\gamma b}$  &  $0.02$  & -       &  $5.04$    &$2.52$  &
                         $0.02$  & -       &  $2.98$    &$1.49$  &
                         $0.03$  & -       &  $3.28$    &$1.64$  &
                         $0.02$  & -       &  $4.32$    &$2.16$  \\
 $\sigma_{\gamma g}^*$&  $0.29$  & $344.37$  &  $6.03$    &$5.14$  &
                         $0.29$  & $172.61$  &  $4.33$    &$3.67$  &
                         $0.37$  & $241.67$  &  $5.87$    &$4.97$  &
                         $0.22$  & $241.67$  &  $4.44$    &$3.78$  \\
 $\sigma_{\gamma c}^*$&  -     & $756.97$  &  $1.34$    &$0.67$  &
                         -     & $439.95$  &  $1.19$    &$0.60$  &
                         -     & $574.42$  &  $1.48$    &$0.74$  &
                         -     & $574.42$  &  $1.08$    &$0.54$  \\
 $\sigma_{\gamma b}^*$&  $0.34$  & -       &  $22.61$   &$11.31$  &
                         $0.34$  & -       &  $13.70$   &$6.85$  &
                         $0.38$  & -       &  $15.06$   &$7.53$  &
                         $0.30$  & -       &  $19.49$   &$9.74$  \\
\hline        Total         &0.67&1188.83& \multicolumn{2}{c|}{58.15}&0.67&660.78& \multicolumn{2}{c|}{37.39}&0.81&880.61& \multicolumn{2}{c|}{44.03}&0.56&880.61& \multicolumn{2}{c}{48.18}  \\
\hline
\end{tabular}
\label{tabmc}
\end{table}
\end{center}
\end{widetext}

We present the cross sections at the $\sqrt{S}=1.30$ TeV LHeC for $m_c = 1.80 \pm 0.10 $ GeV and $m_b = 5.10 \pm 0.20$ GeV in Table~\ref{tabmc}. By adding those two errors in quadrature and summing up all the diquark configurations and production channels together, we obtain
\begin{eqnarray}
&&\sigma_{\langle bb \rangle }^{\rm Total}=0.67^{+0.14}_{-0.11}~\rm{pb},\nonumber \\
&&\sigma_{\langle cc \rangle }^{\rm Total}=880.61^{+308.22}_{-219.83}~\rm{pb},\nonumber \\
&&\sigma_{\langle bc \rangle}^{\rm Total}=46.29^{+12.06}_{-9.18}~\rm{pb}.\nonumber
\end{eqnarray}

\begin{table}[htb]
\caption{Variations for the total cross sections (in unit pb) for the photoproduction of $\Xi_{QQ'}$ at the $\sqrt{S}=1.30$ TeV LHeC under the renormalization/factorization scale set at $\mu=0.75M_T$ and $\mu=1.25M_T$.}
\begin{center}
\begin{tabular}{c| c c c c| c c c c}
\hline - & \multicolumn{4}{c|}{$\mu=0.75M_T$ } & \multicolumn{4}{|c}{$\mu=1.25M_T$ }\\
\hline - & $ \langle bb \rangle_{\textbf{c}}$ & $ \langle cc \rangle_{\textbf{c}}$ & $ \langle bc \rangle_{\bar{\textbf{3}}}$ & $ \langle bc \rangle_{{\textbf{6}}} $& $ \langle bb \rangle_{\textbf{c}}$ & $ \langle cc \rangle_{\textbf{c}}$ & $ \langle bc \rangle_{\bar{\textbf{3}}}$ & $ \langle bc \rangle_{{\textbf{6}}} $\\
 $\sigma_{\gamma g}$  &  $0.03$  & $24.57$  &  $1.79$    &  $1.23$  &
                         $0.02$  & $19.61$   &  $1.39$     & $0.96$  \\
 $\sigma_{\gamma c}$  &  -     & $46.21$  &  $0.29$    & $0.15$&
                         -     & $39.94$   &  $0.25$    &$0.12$  \\
 $\sigma_{\gamma b}$  &$0.02$&-&$2.09$&$1.04$&
                       $0.03$&-&$4.45$&$2.23$  \\
 $\sigma_{\gamma g}^*$&$0.33$&$272.55$&$5.87$&$4.97$&
                       $0.26$&$220.26$&$4.57$&$3.89$  \\
 $\sigma_{\gamma c}^*$&-&$617.91$&$1.38$&$0.69$&
                       -&$534.66$&$1.17$&$0.59$  \\
 $\sigma_{\gamma b}^*$&$0.30$&-&$9.50$&$4.75$&
                       $0.35$&-&$20.24$&$10.12$  \\
\hline        Total         &0.68&961.24& \multicolumn{2}{c|}{33.75}&0.66&814.47&\multicolumn{2}{c}{49.98}  \\
\hline
\end{tabular}
\label{scaleu}
\end{center}
\end{table}

To discuss the scale uncertainty, we set the factorization scale as the renormalization scale, $\mu_f=\mu_r=\mu$. In addition to the scale choice of $\mu=M_T$, we adopt two other scale choices $\mu=0.75M_T$ and $\mu=1.25M_T$ for discussing the scale uncertainty. The scale uncertainty for each diquark configuration and production channel is presented in Table~\ref{scaleu}. The scale uncertainties for the total cross sections are $\sim1\%$, $\sim9\%$, and $\sim27\%$ for $\Xi_{bb}$, $\Xi_{cc}$, and $\Xi_{bc}$, respectively, and we can reduce the scale dependence if we know the ${\beta_i}$-terms of the pQCD series through a next-to-leading order calculation~\cite{pmc3, pmc5}.

\begin{table}[htb]
\caption{Variations for the total cross sections  (in unit pb) for the photoproduction of $\Xi_{QQ'}$ at the $\sqrt{S}=1.30$ TeV LHeC by setting the electron scattering angle cut $\theta_c=64$ and $16$ mrad.}
\begin{center}
\begin{tabular}{c| c c c c| c c c c}
\hline - & \multicolumn{4}{c|}{$\theta_c=16$ mrad} & \multicolumn{4}{|c}{$\theta_c=64$ mrad}\\
\hline - & $ \langle bb \rangle_{\textbf{c}}$ & $ \langle cc \rangle_{\textbf{c}}$ & $ \langle bc \rangle_{\bar{\textbf{3}}}$ & $ \langle bc \rangle_{{\textbf{6}}} $& $ \langle bb \rangle_{\textbf{c}}$ & $ \langle cc \rangle_{\textbf{c}}$ & $ \langle bc \rangle_{\bar{\textbf{3}}}$ & $ \langle bc \rangle_{{\textbf{6}}} $\\
 $\sigma_{\gamma g}$  &  $0.02$  & $20.08$   &  $1.43$    & $0.98$  &
                         $0.03$  & $23.13$   &  $1.66$    & $1.14$  \\
 $\sigma_{\gamma c}$  &  -     & $39.98$   &  $0.25$    & $0.12$&
                         -     & $45.84$   &  $0.29$    &$0.14$  \\
 $\sigma_{\gamma b}$  &  $0.02$  & -       &  $3.56$    &$1.78$&
                         $0.03$  & -       &  $4.13$    &$2.06$  \\
 $\sigma_{\gamma g}^*$&  $0.26$  & $224.83$  &  $4.71$    &$4.00$&
                         $0.31$  & $258.54$  &  $5.47$    &$4.64$  \\
 $\sigma_{\gamma c}^*$&  -     & $535.40$  &  $1.17$    &$0.59$&
                         -     & $613.45$  &  $1.35$    &$0.68$  \\
 $\sigma_{\gamma b}^*$&  $0.31$  & -       &  $16.18$   &$8.09$&
                         $0.36$  & -       &  $18.77$   &$9.39$  \\
\hline        Total         &0.61&820.29& \multicolumn{2}{c|}{42.86}&0.73&940.96& \multicolumn{2}{c}{49.72}  \\
\hline
\end{tabular}
\label{theta2}
\end{center}
\end{table}

As a final remark, we make a discussion on the uncertainties from the electron scattering angle cut $\theta_c$. For the purpose, we set $\theta_c=16$ and $64$ mrad. The results are shown in Table~\ref{theta2}. Within those choices, Table~\ref{theta2} shows the uncertainties from $\theta_c$ are $\sim9\%$ for $\Xi_{bb}$, $\sim7\%$ for $\Xi_{cc}$, and $\sim7\%$ for $\Xi_{bc}$. The small uncertainty from $\theta_c$ shows the WWA is appropriate choice for our calculation.

\section{Summary}\label{sec4}

In the paper, the photoproduction of doubly heavy baryon at the LHeC via the channels $\gamma+g\to \Xi_{QQ'}+\bar{Q}+\bar{Q'}$ and $\gamma+Q\to \Xi_{QQ'}+\bar{Q'}$ has been studied within the framework of NRQCD.

We have found the extrinsic heavy quark mechanism via $\gamma+Q\to \Xi_{QQ'}+\bar{Q'}$ provides a significant production rate comparing to the channel $\gamma+g\to \Xi_{QQ'}+\bar{Q}+\bar{Q'}$, regardless of the suppressions from the heavy quark PDFs. There are four spin-and-color diquark configurations for the production of doubly heavy baryons, i.e., $[^1S_0]_{\bar{\textbf{3}}/{\textbf{6}}}$ and $[^3S_1]_{\bar{\textbf{3}}/{\textbf{6}}}$. It was found that the $[^3S_1]_{\bar{\textbf{3}}}$-diquark state provides dominate contribution to the $\Xi_{QQ'}$ production. While other diquark states can also provide sizable contributions to the $\Xi_{QQ'}$ production, thus a careful discussion of all diquark configurations are helpful for a sound prediction of the doubly heavy baryon production.

The cross sections and its uncertainties caused by different choices of the heavy-quark mass, the renormalization/factorization scale and $\theta_c$ are presented. By taking $m_c=1.80\pm0.10$ GeV and $m_b=5.1\pm0.20$ GeV, we shall have $(6.70^{+1.40}_{-1.10})\times 10^{3}$ $\Xi_{bb}$, $(8.81^{+3.08}_{-2.20})\times 10^{6}$ $\Xi_{cc}$, and $(4.63^{+1.21}_{-0.92})\times 10^{5}$ $\Xi_{bc}$ events to be generated in one operation year at the LHeC under the condition of $\sqrt{S}=1.30$ TeV and ${\cal L}\simeq 10^{33}$ cm$^{-2}$s$^{-1}$, where all diquark configurations and production channels have been summing up. With such amounts of production rate, the LHeC and its updated version, the FCC-$ep$, shall provide a helpful platform for studying the properties of doubly heavy baryon, especially for $\Xi_{cc}$ and $\Xi_{bc}$.

At last, we make a discussion about the light quark component in the doubly heavy baryon $\Xi_{QQ'q}$, where $q$ denotes light quark $u$, $d$, or $s$, respectively. The ratio for a diquark $\langle QQ' \rangle[n]$ evolving to a certain doubly heavy baryon $\Xi_{QQ'u}$, $\Xi_{QQ'd}$, or $\Xi_{QQ's}$ is about $1:1:0.3$~\cite{pythia}. Thus, $43\%$ of $\langle cc \rangle[n]$ are to be fragmented into $\Xi_{cc}^{++}$, $43\%$ for $\Xi_{cc}^{+}$, and $14\%$ for $\Omega_{cc}^{+}$. The same estimation occurs for the production of $\Xi_{bb}^0$, $\Xi_{bb}^-$, and $\Omega_{bb}^-$ or the production of $\Xi_{bc}^+$, $\Xi_{bc}^0$, and $\Omega_{bc}^0$.  \\

{\bf Acknowledgement:} This work was supported in part by the National Natural Science Foundation of China (No.11375171, No.11405173, No.11535002 and No.11625520). The authors would like to thank Xu-Chang Zheng for helpful discussion.

\end{document}